\input harvmac.tex
\input epsf






\Title{}{T-duality of Large N QCD.}

\centerline{$\quad$ { Z. Guralnik}}
\smallskip
\centerline{{\sl University of Pennsylvania}}
\centerline{{\sl Philadelphia PA, 19104}}
\centerline{{\tt guralnik@ovrut.hep.upenn.edu}}

\vskip .3in 

We argue that non-supersymmetric large $N$ QCD compactified on $T^2$ 
exhibits properties 
characteristic of an $SL(2,Z)$ T-duality. 
The kahler structure on which this $SL(2,Z)$ acts is given by 
${m\over N} + i\Lambda^2 A$, where $A$ is the area of the
torus,  $m$ is the 't Hooft magnetic flux on the torus, and $\Lambda^2$ 
is the QCD string tension.


\Date{}

\lref\eguchi{T. Eguchi and H. Kawai, {\it ``Reduction of dynamical 
degrees
		of freedom in the large N gauge theory,''} 
		Phys. Rev. Lett.48 (1982) 1063.}
\lref\nahm{ P. J. Braam and P. van Baal, {\it ``Nahms transformation 
	    for Instantons,''} Commun. Math. Phys. 122 (1989) 267; 
	   P. van Baal, {\it ``Instanton moduli for $T^3 \times R$,''}
	   Nucl. Phys. Proc. Suppl. 49 (1996) 238, hep-th/9512223.}
\lref\verlindetduality{H. Verlinde, talk given at Princeton 1996.}
\lref\dougmoore{M. Douglas and G. Moore, {\it ``D-branes, quivers and 
                ALE instantons,''}hep-th/9603167.}
\lref\EWbstate{E. Witten, {\it ``Bound states of strings and 
p-branes,''}
		Nucl. Phys. B460 (1996) 335, hep-th/9510135.}
\lref\polchbranes{ }
\lref\raminst{ M. Douglas, {\it "Branes within branes,"}
               hep-th/9512077.}
\lref\DBI{ R. G. Leigh, {\it "Dirac-Born-Infeld action from 
             Dirichlet sigma model,"} 
           Mod. Phys. Lett.A4 (1989) 2767.}
\lref\nonlinsusy{ J. Harvey and G. Moore, {\it "On the algebras
		   of BPS states,"} hep-th/9609017}
\lref\verlindehac{ H. Verlinde and F. Hacquebord, 
		  {\it "Duality symmetry of $N=4$ Yang-Mills
			theory on $T^3$,"} hep-th/9707179.}
\lref\zbranes{ Z. Guralnik and S. Ramgoolam, { \it "From
		0-Branes to Torons,"} hep-th/9708089.}
\lref\bfss{ T. Banks, W. Fischler, S. Shenker, L. Susskind,
		{\it "M Theory as a matrix model: a conjecture,"}
		 Phys. Rev. D55 (1997) 5112-5128, hep-th/9610043.}
\lref\lcone{ L. Susskind, {\it "Another conjecture about M(atrix)
		theory,"} hep-th/9704080.}
\lref\blau{ M. Blau and M. O'Loughlin, {\it "Aspects of U-duality
		in Matrix Theory,"} hep-th/9712047.}
\lref\hull{ C. Hull, {\it "U-duality and BPS spectrum of Super 
		Yang-Mills theory and M theory,"} hep-th/9712075.}
\lref\joao{ Private communication }
\lref\rabinov{ N. A. Obers, B. Pioline, E. Rabibovici, {\it 
		"M theory and U-duality on $T^d$ with gauge 
		backgrounds,"} hep-th/9712084.}
\lref\noncom{ A. Connes, M. Douglas, A. Schwarz, 
	{\it "Noncommutative geometry and Matrix theory:
	 compactification on tori,"} hep-th/9711162.}
\lref\hulldoug{ M. Douglas and C. Hull, {\it "D-branes and
		the noncommutative torus,"} hep-th/9711165.}
\lref\thooft{ G. 't Hooft, {\it "A property of electric and
		magnetic flux in non-abelian gauge theories,"}
	       Nucl. Phys. B153 (1979) 141-160.}
\lref\rudd{ R. Rudd, {\it "The string partition function
		for QCD on the torus,"} hep-th/9407176.}
\lref\gtaylor{ D. Gross and W. Taylor, {\it 
		"Two dimensional QCD is a string
		theory,"} Nucl. Phys. B400 (1993) 181,
		hep-th/9301068.}
\lref\gross{ D. Gross, {\it "Two dimensional QCD as a string
		theory,"} Nucl. Phys. B400 (1993) 161,
		hep-th/9212149.}
\lref\polyakov{A. Polyakov, {\it ``String representations and hidden
		symmetries for gauge fields,''} 
		Phys. Lett. 82B (1979) 247-250.
                {\it ``Gauge fields as rings of glue,''}
		Nucl. Phys.B164 (1980) 171-188.} 
\lref\igor{K. Demeterfi, I. Klebanov,  Gyan Bhanot, 
	   {\it ``Glueball spectrum in a $(1+1)$ dimensional 
	     model for QCD,''} Nucl. Phys.B418 (1994) 15-29, 
hep-th/9311015.}
\lref\daleykleb{S. Dalley, I. Klebanov, {\it String spectrum of 
$(1+1)$
	        dimensional large $N$ QCD with adjoint matter,''}
	        Phys. Rev. D47 (1993) 2517-2527.}
\lref\antonuccio{F. Antonuccio and S. Dalley, {\it `` Glueballs
		 from (1+1) dimensional gauge theories with 
		transverse degrees of freedom,''} 
                 Nucl. Phys. B461 (1996) 275-304, hep-ph/9506456.}
\lref\migdal{A. Migdal, {\it ``Recursion equations in gauge 
theories,''}
             Sov.Phys.JETP42 (1975) 413, Zh.Eksp.Teor.Fiz.69 (1975) 
810-822.}
\lref\witten{E. Witten, {\it ``Two dimensional gauge theories 
revisited.''}
	      J. Geom. Phys. 9 (1992) 303-368, hep-th/9204083.}
\lref\taylor{W. Taylor, IV, {\it ``D-brane field theory on 
		compact spaces,''} Phys. Lett. B394 (1997) 283-287,
		hep-th/9611042.}
\lref\sav{S. Sethi and L. Susskind, {\it ``Rotational invariance in 
	   the M(atrix) formulation of IIB theory,''} 
          Phys. Lett. B400 (1997) 265-268, hep-th/9702101.}
\lref\torons{Z. Guralnik and S. Ramgoolam, 
		{\it ``Torons and D-brane bound states,''}
	     Nucl. Phys. B499 (1997) 241-252, hep-th/9702099.}
\lref\tdlty{A. Giveon, M. Porrati, and E. Rabinovici, 
	{\it ``Target space duality in string theory,''}
 	 Phys. Rept. 244 (1994) 77-202, hep-th/9401139.} 
\lref\tseytlin{   }
\lref\blautom{M. Blau and G. Thompson, {\it ``Lectures on 2-D gauge 
	theories: topological aspects and path integral techniques,''}
	Trieste HEP Cosmol. 1993: 175-244, hep-th/9310144.}
\lref\ramoore{S. Cordes, G. Moore, S. Ramgoolam, 
	{\it ``Large N 2-D Yang-Mills
 	theory and topological string theory,''} 
	Commun. Math. Phys.185 (1997) 543-619, hep-th/9402107.} 
\lref\horava{P. Horava, {\it ``Topological rigid string theory and 
two-dimensional QCD,''}\ \ \
Nucl. Phys. B463 (1996) 238-286, hep-th/9507060.} 
\lref\dougjev{M. Douglas, {\it ``Conformal field theory techniques 
	in large N Yang-Mills theory,''} RU-93-57, Presented at 
Cargese
	Workshop on Strings, Conformal Models and Topological Field 
Theories, 
	Cargese, France, May 12-26, 1993, hep-th/9311130.}
\lref\mukai{S. Mukai, Invent. Math.7 (1984) 189. }
\lref\nahmhimself{W. Nahm, {\it ``The construction of all self-dual 
	multimonopoles by the ADHM method,'' } 
	Phys. Lett.90B (1980) 413.}	
\lref\morepoly{A. Polyakov, {\it ``String theory and quark 
confinement,''}
		hep-th/9711002. 
		{\it ``Confining strings,''} Nucl.Phys.B486 
		(1997) 23-33, hep-th/9607049. }
\lref\rusakov{B. Rusakov, {\it ``Loop averages and partition functions
		in U(N) gauge theory on two-dimensional manifolds,''}
		Mod.Phys.Lett.A5 (1990) 693-703.}
\lref\gtr{O. Ganor, W. Taylor, and S. Ramgoolam, {\it ``Branes,
		fluxes and duality in M(atrix) theory,''}
		Nucl.Phys.B492 (1997) 191-204, hep-th/9611202.} 
\lref\thooftor{G. 't Hooft, {\it ``Some twisted selfdual solutions 
for the
	Yang-Mills equations on a hypertorus,''} 
	Commun.Math.Phys.81 (1981) 267-275.} 
\lref\decon{B. Svetitsky and L. Yaffe, {\it ``Critical behavior at
	finite temperature confinement transitions,''} 
	Nucl.Phys.B210 (1982) 423.}
\lref\maldads{J. Maldacena, {\it ``The large N limit of superconformal 
field theories and supergravity,''} e-Print Archive: hep-th/9711200.} 
\lref\adskleb{S. Gubser, I. Klebanov, and 
A. Polyakov, {\it ``Gauge theory
correlators from noncritical string theory,''} 
e-Print Archive: hep-th/9802109.} 
\lref\ads{E. Witten, {\it ``Anti-de-Sitter space, thermal phase
transition,  and confinement in gauge theories,''}
e-Print Archive: hep-th/9803131. } 
\lref\mqcd{E. Witten, {\it ``Branes and the dynamics of QCD,''}
Nucl.Phys.B507 (1997) 658-690, 
e-Print Archive: hep-th/9706109.}
\lref\oog{K. Hori and H. Ooguri, {\it ``Strong coupling dynamics 
	of four-dimensional N=1 gauge 
	theories from M theory five-brane.''} 
	Adv.Theor.Math.Phys.1 (1998) 1-52,  hep-th/9706082.}
\lref\moi{Z. Guralnik, {\it ``Strings and discrete fluxes of QCD,''}
	in preparation.}
\lref\dougkazak{M. Douglas and V. Kazakov, {\it ``Large N phase 
transition in continuum QCD in two-dimensions,''} 
Phys. Lett.B319 (1993) 219-230,
e-Print Archive: hep-th/9305047} 
\lref\holom{ E. Witten, {\it ``Anti-de Sitter space and holography,''}
e-Print Archive: hep-th/9802150 } 
\lref\thooftgen{???}  
\lref\me{Z. Guralnik, {\it ``Duality of large N Yang Mills theory 
on $T^2 \times R^2$,''}
e-print Archive: hep-th/9804057 }
\lref\largN{ G. 't Hooft, {\it ``A planar diagram theory for strong 
interactions,''} 
Nucl. Phys.B72 (1974) 461.} 
\lref\mqcd{ E. Witten, {\it ``Branes and the dynamics of QCD,''}
Nucl. Phys.B507 (1997) 658-690, e-print Archive: hep-th/9706109.} 

\newsec{ Introduction } 

Following 't Hooft's discovery that the large N expansion of QCD is 
an expansion in
the  genus of feynman graphs, 
\largN, 
it has been suspected that large $N$ QCD is a string theory in 
which the string coupling
is given by $1\over N$. 
This correspondance is best  understood for  pure 
$QCD_2$ \gross\gtaylor\ramoore\horava.  
The QCD string in higher dimensions has yet to be constructed,  
although much progress has been made recently based on the Maldacena 
conjecture \morepoly\adskleb\maldads\holom\ads. 
Assuming the existance of a string description,  large $N$ QCD will
inherit any self-dualities of this description.  In this talk
we address the question of whether the QCD string,  when 
compactified on a two torus,  has a self T-duality.  
Such a T-duality would be generated by 
\eqn\gener{\eqalign{\tau\rightarrow\tau+1\cr
			\tau\rightarrow -\bar\tau\cr
			\tau\rightarrow -{1\over\tau}}}
for $\tau = B + i \Lambda^2 A$,  where $A$ is the area of the torus, 
$B$ is a two form modulus, and $\Lambda^2$ is the string tension. 
For simplicity we consider only square tori.
We shall argue
that the two form modulus is
${m\over N}$ where $m$ is the 't Hooft magnetic 
flux \thooft\ through the torus.  
This quantity has the desired properties of periodicity,   
and continuity in the large $N$ limit.   
We suspect that pure QCD is not exactly self T-dual,  although 
another theory in the same universality class may be self dual.
In two dimensions the  partition function of $QCD_2$ is invariant 
under T-duality
after a simple modification \rudd\me.  In the case of $QCD_4$ on 
$T^2\times R^2$, 
there are qualitative properties consistent with self T-duality.  
If $QCD_4$ on $T^2 \times R^2$ is T-dual to a another theory
in the same universality class,  there may be  computationally useful 
consequences.  By dualizing pure $QCD_4$ on a very large torus,  
one would  obtain 
a $QCD_2$-like theory with two adjoint scalars.
Such theories have been used as toy models which mimic some of
the dynamics of pure $QCD_4$\daleykleb\igor\antonuccio.  
However these models lack 
a $U(1)\times U(1)$ symmetry which could generate two extra dimensions. 
We shall propose a model which does have such a symmetry
in the large $N$ limit. This model is obtained by a dimensional 
reduction of $QCD_4$ preserving the $Z_N \times Z_N$ global symmetry 
generated by large gauge transformations on the torus.     

\newsec{$QCD_2$ and T-duality}

Consider pure Euclidean $QCD_2$ on a two-torus. 
The partition function for vanishing 't Hooft flux is 
given by \rusakov\migdal 
\eqn\twod{Z = \sum_R e^{g^2AC_2(R)},}
where $C_2(R)$ is the quadratic casimir in the representation $R$.
When the 't Hooft flux $m$ is non-vanishing,  the partition 
function is given
by \witten\me\
\eqn\twodtwist{Z = \sum_R e^{g^2AC_2(R)}{\Tr_R(D_m)\over d_R},}
where $d_R$ is the dimension of the representation.
$\Tr_R (D_m)$ is the trace of the 
element in the center of $SU(N)$ corresponding to the 't Hooft flux.
For instance in a representation for which the Young Tableaux 
has $n_R$ boxes,  
\eqn\twistrep{D_m = e^{2 \pi i {m\over N} n_R}.}
Evaluating the free energy in the planar $N\rightarrow\infty$ 
limit, as done
in \gtaylor\ for vanishing $m$,  gives 
\eqn\zmodular{F = ln  \left| {e^{2 \pi i {\tau\over 24}} 
		\over \eta(\tau)} \right| ^2}
where $\eta$ is a Dedekind eta function,  and 
\eqn\modparam{\tau = {m\over N} - {\lambda A \over 2\pi i},}
with $g^2N = \lambda$.  This is not quite invariant under \gener.  
However
the modified free energy
\eqn\modified{{\cal F} = F + {1\over 24}\lambda A - {1\over 2} 
ln (\lambda A)}
is  modular invariant.  The additional term proportional 
to $A$ is a local 
counterterm.  However it is not clear what modification 
of the action could  
account for the term $ln (\lambda A)$.  Nonetheless this 
term is very simple,
so it seems that pure $QCD_2$ is almost self T-dual.

\newsec{'t Hooft flux and two form moduli}

We have seen that under T-duality, the 't Hooft flux $m/N$ 
behaves like a two 
form modulus of a string description.  There are several 
reasons one could have 
guessed
this correspondance between magnetic flux and the string theory two form.
First, $m/N$ is periodic and continuous (for $N\rightarrow\infty$). 
Second,  in $QCD_2$  with  $\lambda A \rightarrow 0$,  the partition 
function is an integral over the moduli space of flat connections.  
The dimension
of this space is invariant under $SL(2,Z)$ transformations acting on 
the doublet (m, N) \zbranes. 
Under these transformations  $m/N$ transforms precisely like the 
Kahler structure $\tau$
of string theory,  which for vanishing area is just the 
two form modulus. 

Finally using the correspondance between D-branes and 
Yang Mills theories one can
argue that under the appropriate conditions the magnetic 
flux is equal to the NS-NS
two-form modulus.   Consider a system of parallel 
D-branes stretched between NS5-branes which is described by a theory with
a mass gap,  such as pure $SU(N)$  ${\cal{N}} =1$ Yang-Mills \mqcd\oog. 
Since parallel D-branes generally give rise to
a $U(N)$ bundle,  let us construct a $U(N)$ bundle in which the $U(1)$
trace degree of freedom is frozen, so that there  are no masseless $U(1)$ 
degrees of freedom.   
On a torus the fields
are periodic up to $U(N)$ gauge 
transformations, $U_1$ and $U_2$,  which for a 
$U(N)$ bundle satisfy 
\eqn\bundle{U_1U_2U_1^{\dagger}U_2^{\dagger} = I.}
The $U$'s may be written as products of $U(1)$ 
and $SU(N)$ pieces,  so that
the above equation becomes
\eqn\prodcut{e^{i\int_{T^2}F_{12}}e^{-2\pi i {m\over N}} = I,}
where $F_{\mu\nu}$ is the $U(1)$ field strength. 
If the $U(1)$ degree of freedom is frozen,  the locally gauge invariant
combination $F_{\mu\nu} - B_{\mu\nu}$  must vanish.  Thus one obtains
\eqn\obtns{{m\over N} = \int_{T^2}B.}
Note this relation was obtained for finite N.  
We shall argue elsewhere \moi\ that 
a periodic potential is generated for $B$ when there is a mass gap. 
In this case one can not obtain 
a continuous class of theories on non-commutative tori \noncom\hulldoug\ 
by varying $B$.  
Having established \obtns,  one must still show that $B$ behaves like
a two-form modulus of the QCD-string as well as the IIA string.  
This means that $B$ should be the imaginary part of the action of a QCD 
string wrapping the torus.   This can be argued \moi\ by
lifting the brane configuration to $M$ 
theory,  in which case the IIA string and the QCD-string 
are homotopic \mqcd.
 
Note that if a QCD T-duality exists,  it is not the same T-duality in
the IIA theory for a variety of reasons.  
It can 
only exist in the $N\rightarrow\infty$ limit, when ${m\over N}$ becomes
a continuous parameter.  
Also, there is no D2-brane charge in the brane costruction of QCD,  so
that after a IIA T-duality the D4-brane charge vanishes. Furthermore 
the Kahler 
structure of the QCD-string is different 
from that of the IIA string because
the string tensions are different.

\newsec{$QCD_4$ and T-duality}

The four dimensional QCD string is much less understood than the 
two dimensional QCD string.  However
there are qualitative properties of large N 
$QCD_4$ on $T^2 \times R^2$ which are
consistent with self T-duality.  We take the time 
direction to lie in $R^2$.
This theory has a $U(1) \times U(1)$ translation
symmetry on the torus.  If the theory is self T-dual it must  
have another 
$U(1) \times U(1)$ symmetry corresponding 
to translations on the dual torus.
Large gauge transformations on the torus generate a global 
$Z_N \times Z_N$ symmetry,  which becomes 
continuous as $N\rightarrow\infty$.
If this  symmetry is a translation
symmetry on a dual torus,  then eigenstates of this symmetry 
should have energies proportional to $1/R_d$, where
$R_d$ is the radius of the small dual torus.  This is consistent 
with electric 
confinement.  The eigenstates of $Z_N \times Z_N$ transformations carry 
electric flux \thooft,  which have energy 
proportional to $R= 1/(\Lambda^2 R_d)$, 
where $R$  is the radius of the original torus.  
(We are considering the case with vanishing magnetic flux).

If the magnetic flux is non-vanishing,  
then $\tau \rightarrow -{1\over \tau}$ 
does not invert the area of the torus.  Thus it 
would not make sense in this case
to exchange the winding number of the QCD string,  
or the electric flux,  with
the QCD momentum.  In string theory the momentum that gets 
exchanged with winding
number under 
$\tau \rightarrow -{1\over \tau}$  is $P_i = p_i - B_{ij}w^j$.
Here $p_i$ is the velocity of the string, and $w^j$
is the winding number:  $X^i(t,\sigma) = p^i t + w^i\sigma + \dots$.  
Under $\tau\rightarrow\tau + 1$,  $p_i$ and $w^j$
are invariant but $P_i$ is shifted.   Therefore the quantity 
in QCD corresponding to 
$P_i$ is $p_i - {m\over N} \epsilon_{ij}e^j$,  where $p_i$ is the 
usual momentum and $e^i$ is the electric flux.  
The term ${m\over N} \epsilon_{ij}e^j$ has precisely the 
form  of a cross product of 
electric and magnetic fluxes and can be thought of as the 
contribution of 't Hooft
fluxes to the momentum.

\newsec{Why there might not be T-duality}

If this $Z_N \times Z_N$ symmetry is spontaneously 
broken for a sufficiently small
torus then QCD can not be self T-dual.  Shrinking one cycle of the
torus while keeping the other fixed would break 
the $Z_N$ associated with the small torus,
just as in a finite temperature deconfinement transition. 
We do not know what happens if both cycles of the torus 
are shrunk simultaneously,
but we can not rule out the possibility that 
both $Z_N$'s are spontaneously broken. 
Even so there may still be some theory in the QCD universality class
for which the $Z_N \times Z_N$ symmetry is never spontaneously broken. 
Recent arguments suggest that large N $QCD_4$ can be 
described by a critical string
theory in a five dimensional background,  whose boundary 
is the QCD world 
volume\morepoly\adskleb\maldads\holom\ads. 
In this picture,  spontaneous breaking of $Z_N \times Z_N$ 
would require a phase 
transition to a five dimensional target space geometry in 
which both cycles of the 
boundary torus are contractible\ads.  Perhaps such a 
transition does not exist. 

\newsec{$QCD_4$ from two dimensions}

If large N QCD on $T^2 \times R^2$ were self T-dual,  
pure QCD on $R^4$ would be dual to 
QCD on $R^2$ with two adjoint scalars.
In fact such a model has been used to approximate the 
dynamics of pure QCD in $4$ 
dimensions  \daleykleb\igor\antonuccio.  
The adjoint scalars in this model 
play the role of transversely polarized gluons.   
In \antonuccio\ the spectrum of this
two dimensional model was  computed by discrete 
light cone quantization and
compared to the glueball spectrum of pure 4-d QCD computed using
Monte-Carlo simulation.  The degree of numerical accuracy allows only
crude comparison,  however the spectra have some qualitatively 
agreement.  Perhaps in the $N\rightarrow\infty$ limit the 
agreement is more than 
just qualitative.  However the usual models with adjoint 
scalars are incomplete
since they lack a $Z_N \times Z_N$ symmetry.   
A more careful dimensional
reduction would give a non-linear sigma model of the form
\eqn\goodaction{S_{SU(N)} =  
	{N\over\lambda_{2d}}\int d^2 x Tr \left( F_{\mu\nu}^2 
	+ {1\over R_s^2} (h_i D_{\mu} h_i^{\dagger})^2 + 
	{1\over R_s^4} [h_2,h_3][h_2^{\dagger},h_3^{\dagger}]\right).}
where $R_s$ is the radius of the cycles 
of the torus, and $h_i$ an element of
$SU(N)$.  Writing $h_i = \exp(iR_sX^i)$ and taking $R_s \rightarrow 0$ 
with $X^i$ fixed gives the usual naive 
reduced action.  Note that the naive  
reduced action requires a mass counterterm 
for $X^i$.  In terms of the $h_i$ fields,
such a term would look like ${1\over R_s^2} \sum_i Tr h_i$,   
which is prohibited by 
$Z_N \times Z_N$ symmetry.   Of course 
the $Z_N \times Z_N$ symmetry might be 
spontaneously broken for sufficiently small $R_s$, 
in which case the large N limit can
not generate extra dimensions.  At tree level there are flat directions 
which connect vacua related by the $Z_N$ symmetry.  
If these flat directions 
are not lifted quantum mechanically, as in a supersymmetric 
version of this theory,
then long range fluctuations in two dimensions
would prevent spontaneous symmetry breaking at finite $N$.  
Unfortunately this last statement is not always true in the 
large $N$ limit.

\newsec{Conclusion}

We have shown that large $N$ QCD has properties consistent with
the existence of self T-duality when compactified on a torus.   
It may be that QCD is
not really self T-dual,  but that some QCD-like theory is. 

\bigbreak\bigskip\bigskip

{\bf Acknowledgments}\nobreak

I am thankful to Antal Jevicki, Igor Klebanov, Joao Nunez, Burt Ovrut,    
Sanjaye Ramgoolam,  Washington Taylor,  and Edward Witten 
for enlightening conversations. 

\listrefs

\end